# The effect of the second nearest neighbor interaction on the population transfer in a four-particle Landau-Zener system

Aarash Maroufian, Mehdi Hosseini*

*Department of Physics, Shiraz University of Technology, Shiraz, Iran(*hosseini@sutech.ac.ir)*

Population transfer in quantum systems has always been an interesting area in physics since the introduction of quantum mechanics. In this paper, transition probabilities for a coupled system consisting of four two-level particles are studied by solving Landau – Zener Hamiltonian. The effects of the first and second nearest neighbors' interactions are investigated. Presented results indicate that the second nearest neighbors' interactions will decrease the transition probability when the coupling strength for each neighborhood has the same sign. The fast sweep effect on transition probability is also studied here.

## I. Introduction

Several methods have been introduced to study population transfer in a two-level system. In this area, Landau-Zener model has been used vastly [1-3]. This model was first introduced separately by Landau and Zener [4,5]. Due to its mathematical simplicity, it is used in quantum data computation [6-9], quantum dots' excitation [10], atomic clocks [11, 12], quantum optics [13, 14], controlling chemical reactions [15], and quantum computing [16 – 18]. This model is also a useful method to study quantum systems with more than two levels [19].

Other models are also introduced to study the population transfer in a two-level system as well. Using a  -pulse laser for population transfer in a quantum system causes a fast and complete transition. However, this transition is not sustainable. High dependency of transition accuracy to laser's intensity is the other downside to this model [20, 21]. STIRAP (Stimulated Raman

Adiabatic Passage) [22 - 24] method is a semi-classical or quantum model, introduced to apply population transfer by mean of a two-photon process and require an intermediate state. This method is vastly used in atomic and molecular physics [21, 22, 25, 26].

Using chirped lasers [27, 28] is another common fully or semi quantum method used to control population transfer. Chirped laser pulse simplifies the experimental limitations of Landau-Zener model by removing the essentiality of assuming infinite times for the transition process [27, 29]. In this method, an ultrashort laser pulse is tuned up to terawatt levels.

Landau-Zener model is a semi-classical method which is based on a time-dependent Hamiltonian, giving the probability of a non-adiabatic transition between two states [30]. Meanwhile, the adiabatic Landau-Zener model is a well-known model to study population transfer, where the changes in the system's energy are smooth [22, 31 - 33].

The interaction between spins of an electron and an external magnetic field is known as spin-orbit interaction, connecting external magnetic field and magnetic momentum of an electron through Zeeman's term ($-\mu_B.B$) [34], a common method to apply the population transfer in an electron is accomplished by using a magnetic field [35]. The experimental setup is achieved by using sets of magnets [36].

In this paper, population transfer for a two-state four-particle coupled system is studied by the adiabatic Landau- Zener model. Effects of changes in the magnetic field and second neighbors' interactions are considered to assume the mathematical relation between interactions.

## II. Theory

Transition probabilities for a two-level system with four particles are calculated by solving Schrodinger's equation. The Hamiltonian is calculated by the summation of free particles Hamiltonian in a four-particle system and interaction Hamiltonians.

The free four particles Hamiltonian is simply given from single-particle LZ Hamiltonian [37, 38]:

$$H_{SPLZ} = \begin{pmatrix} \Gamma \frac{t}{2} & g \\ g & -\Gamma \frac{t}{2} \end{pmatrix}, \tag{1}$$

where $t$ is the magnetic field's energy, while $\Gamma$ has a linear relation with the magnetic field and $g$ is the tunneling energy between states when the system is in its initial condition at $t = -\infty$.

In this paper, the interactions between two neighbors are represented by two separate Hamiltonians to illustrate the interactions between nearest and second closest neighborhoods. Therefore, the total Hamiltonian is represented as:

$$H = H_{int,fnn} + H_{int,snn} + H_{FPLZ}. \tag{2}$$

$H_{FPLZ}$ is the four-particle LZ Hamiltonian. $H_{int,fnn}$ and $H_{int,snn}$ are the first and second nearest neighbors Hamiltonians are as follows [39].

$$H_{int,fnn} = J_1(\sigma_z \otimes \sigma_z \otimes I_{2\times2} \otimes I_{2\times2} + \sigma_z \otimes I_{2\times2} \otimes I_{2\times2} \otimes \sigma_z + I_{2\times2} \otimes I_{2\times2} \otimes \sigma_z \otimes \sigma_z + I_{2\times2} \otimes \sigma_z \otimes \sigma_z \otimes I_{2\times2})$$

$$H_{int,snn} = J_2(\sigma_z \otimes I_{2\times2} \otimes \sigma_z \otimes I_{2\times2} + I_{2\times2} \otimes \sigma_z \otimes I_{2\times2} \otimes \sigma_z), \tag{3}$$

in which *J* is coupling strength between particles, *I* is the unit matrix, $\sigma_z$ is *z* component of Pauli matrix [40, 41].

Fig. 1 illustrates the idea of first and second nearest neighbors in a four-particle system, where particle 1 is in the first neighborhood with particles 2 and 3, while particle 4 is considered as its second neighbor. The values and ratio of $J_1$ and $J_2$ are different for various materials, although the interactions between second neighbors are usually smaller than the interactions between first neighbors [42].

The calculations are done with dimensionless parameters. The capped parameters are defined as dimensionless parameters, where $\hat{g} = 1, \hat{J} = J/g, \hat{\Gamma} = \Gamma \hbar^2/g^2$ and $\hat{t} = t/\ddagger$, where $\ddagger = \hbar/g$.

### III. Result and discussion

Fig. 2 shows the transition probabilities for one particle of the system versus time. In this figure, $\hat{\Gamma} = 1$, while coupling strength values are assumed to be as described in the figure, where $\hat{J}_1$ demonstrates coupling strength for the first neighbors and $\hat{J}_2$ represents the coupling strength of the second nearest neighbors. Negative signs of coupling strength in nearest neighborhoods demonstrate ferromagnetic behavior of the system, while positive signs identify an antiferromagnetic system. The time range is chosen between -30 and 30, so a stable transition could be observed. As mentioned before, the system is in its ground states at $\hat{t} = -\infty$, becoming stable while transferred to the excited state at $\hat{t} = -30$ and $\hat{t} = 30$. W. Final transition probabilities are used to observe the final state of the system. When both coupling strength values are positive, a complete transition is observable. The transition probability is close to 1 when one of the values of coupling strength is positive. However, when both coupling strength values are negative, the transition probability is reduced to 0.5. In slow sweep mode (this case), transition probabilities are dependent on coupling strength values and not on the magnetic field energy.

Fig 3 shows the transition probabilities for $\hat{\Gamma}$=7 to observe the effect of increasing the magnetic energy sweeping rate [43, 44]. The coupling strength values are (a): $\hat{J}_1 = 1$, $\hat{J}_2 = 1$, (b): $\hat{J}_1 = 1$, $\hat{J}_2 = -1$, (c): $\hat{J}_1 = -1$, $\hat{J}_2 = 1$, (d): $\hat{J}_1 = -1$, $\hat{J}_2 = -1$. When coupling strength values in both neighborhoods have positive signs (perfect antiferromagnetic), similar to the previous case, the complete transition is observed. When coupling strength values of both neighborhoods are negative (perfect ferromagnetic), transition probabilities are dropped to 50%. As could be seen, transition probabilities are dependent on the values of magnetic field energy, as well as coupling strength values. This figure also illustrates that the transition probability reaches a stable state more quickly for bigger values of $\hat{\Gamma}$.

Final Transition Probabilities of Excited states (FTPE) is defined as following for further investigations. This parameter is defined as the average value of 10% of the final values of transition probabilities for excited states with respect to time, which the mathematical approach is given in (Eq. 4):

$$FTPE = \frac{1}{0.1T}\int_{0.9T}^{T} E.S(t)dt \qquad (4)$$

where *E.S*(t) represents the magnitude of probability for Excited States (E.S) as a function of normalized time and *T* is the total time.

Fig. 4 shows the FTPE versus normalized coupling strength and normalized magnetic field's energy rate when the second neighbor's interactions are not included ($\hat{J}_2 = 0$). Values of normalized coupling strength change between -2 and 2 for nearest neighbors, where negative values of coupling strength demonstrate ferromagnetic properties and positive values define antiferromagnetic properties of the system. Values of $\hat{r}$ changes between 0.2 and 2 (slow sweep range). Based on the results that was presented in figure 2 and 3, FTPE values will be smaller for higher values of $\hat{r}$. This figure illustrated that for positive coupling strengths the transition is complete but for negative coupling strengths, FTPE is about 0.6. Furthermore, the magnetic field's energy rate does not change the FTPE effectively in the studied range. The results for this system are comparable to the case that a three-particle system was studied before [45] and it could be concluded that the effect of the number of particles on FTPE in few-body systems in absence of long-range interaction is not significant.

The effects of the next nearest neighbor are investigated in Fig. 5. Fig 5-a shows the results for FTPE versus coupling strength for first and second nearest neighbors and different magnetic field energies. Here, $\hat{r}=1$, while as before, coupling strength value alters between -2 and 2. Fig 5-b and 5-c represent FTPE while the magnitude of magnetic field energy is fixed at 7 and 13.8. As could be seen in these figures, when the magnetic field energy magnitude is increased, the FTPE values drop. The areas with complete transition shrink as the magnitude of $\hat{r}$ increases. That means complete transition can occur in bigger values of coupling strength by the increase in $\hat{r}$ magnitude.

Three diagrams are presented in Figure 5, in which each one has four regions. In this figure, FTPE is studied, assuming the system is changing between ferromagnetic and antiferromagnetic behavior (change in the sign of first neighborhoods' coupling strength values). In this figure, the transition probability value is maximum, while both coupling strength values are positive. On the other hand, when both coupling strength values are negative, the transition probability is at its smallest value. In regions that coupling strength values have opposite signs, a phase transition could be observed. In these cases, a mixed state is observed. These mixed states tend to change complete transition while the diagrams are shifting to the regions that both coupling strength values have similar signs. These regions with complete transitions shrink as the value of $\hat{r}$ increases.

## IV. Conclusion

In this paper, transition probabilities for a four-particle two-level system are studied by calculating Final Transition Probabilities of Exited states (FTPE) in different ranges of $\hat{r}$.

Initially, second neighbors' interactions are neglected. The result was close to the case in which FTPE was calculated in a three-particle system. There is only a little difference in FTPE for negative values of coupling strength. This might be occurred because of the low ratios of the second neighbors' interaction parameter.

FTPE values also were calculated when the interactions of second nearest neighbors were included. Coupling strength values for both neighborhoods change independently between -2 and 2, and four regions are observed in FTPE graphs. FTPE values are maximum when signs of coupling strength in both neighborhoods are positive. On the other hand, when coupling strength values for both neighborhoods are negative, FTPE value is at its lowest. A mixed state is observed for FTPE in other ranges (opposite signs of coupling strength for each neighborhood).

## Acknowledgement

Special thanks to Fatemeh Ahmadinouri for helpful comments.

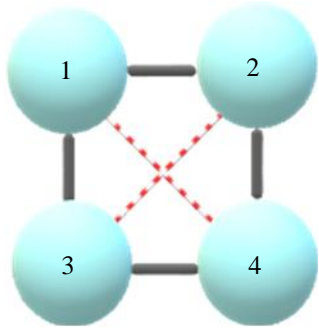

Fig 1: periodic boundary condition in a four particle two-level particle system, particles, 2 and 3 are closest neighbors with respect to particle 1, where particle 4 (on the diameter) is in the second closet neighborhood from particle 1 (connected to particle 1 through the dotted line).

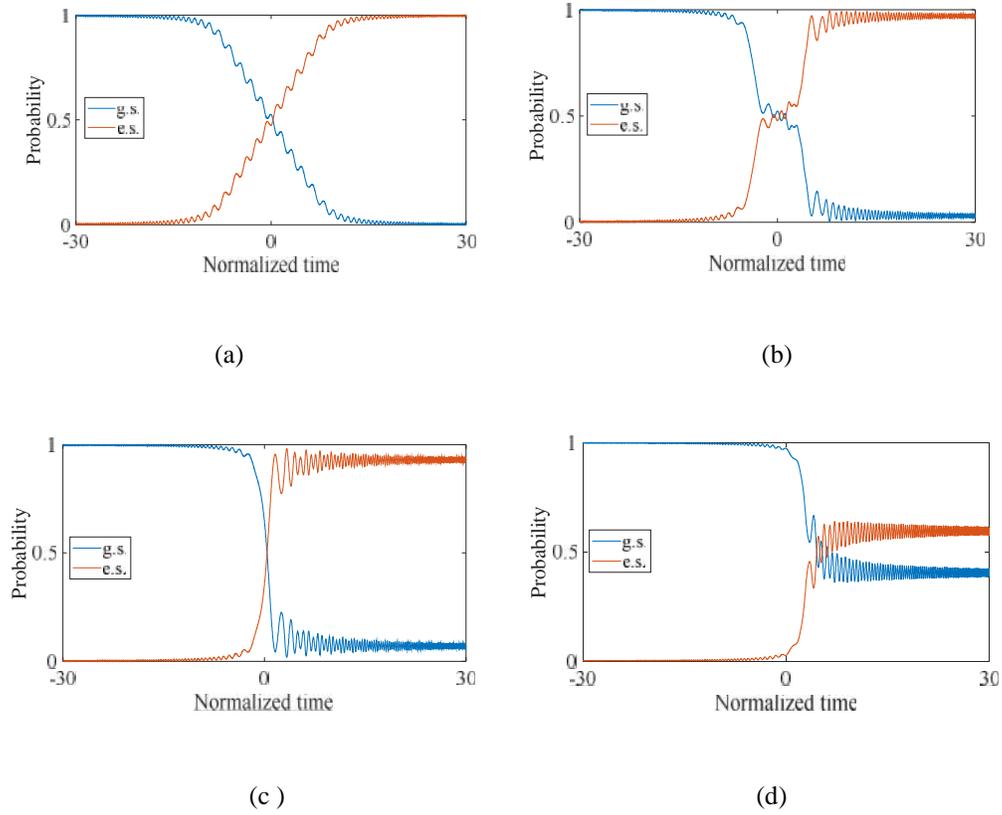

Fig 2: transition probabilities with respect to time, while Alpha = 1 for

(a): $\hat{J}_1 = 1$, $\hat{J}_2 = 1$, (b): $\hat{J}_1 = 1$, $\hat{J}_2 = -1$, (c): $\hat{J}_1 = -1$, $\hat{J}_2 = 1$, (d): $\hat{J}_1 = -1$, $\hat{J}_2 = -1$

(Orange graph represents excited states, while the blue graph represents ground states.)

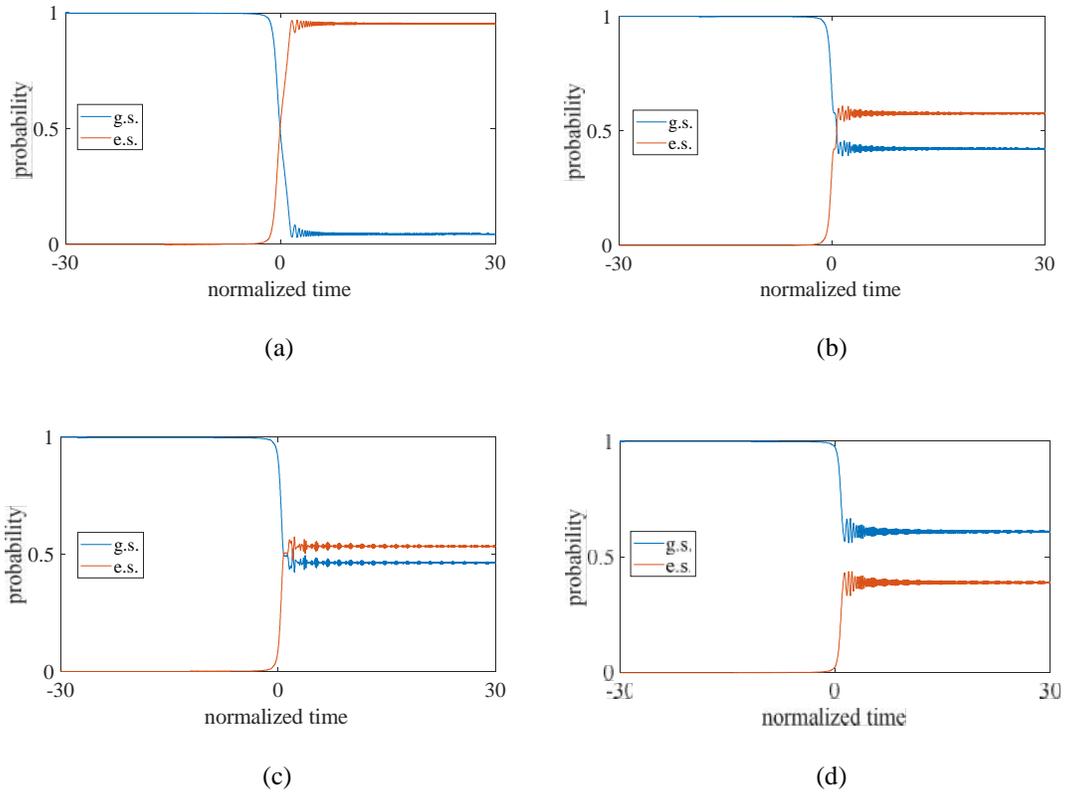

Fig 3: transition probabilities with respect to time, while Alpha = 7 for

(a): $\hat{J}_1 = 1$, $\hat{J}_2 = 1$, (b): $\hat{J}_1 = 1$, $\hat{J}_2 = -1$, (c): $\hat{J}_1 = -1$, $\hat{J}_2 = 1$, (d): $\hat{J}_1 = -1$, $\hat{J}_2 = -1$

(Orange graph represents excited states, while the blue graph represents ground states.)

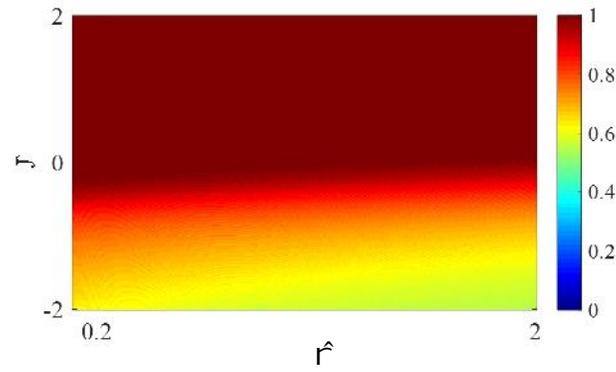

Fig 4: FTPE verses $\hat{J}$ and $\hat{r}$ for a four particle system while second neighbor interactions are neglected.

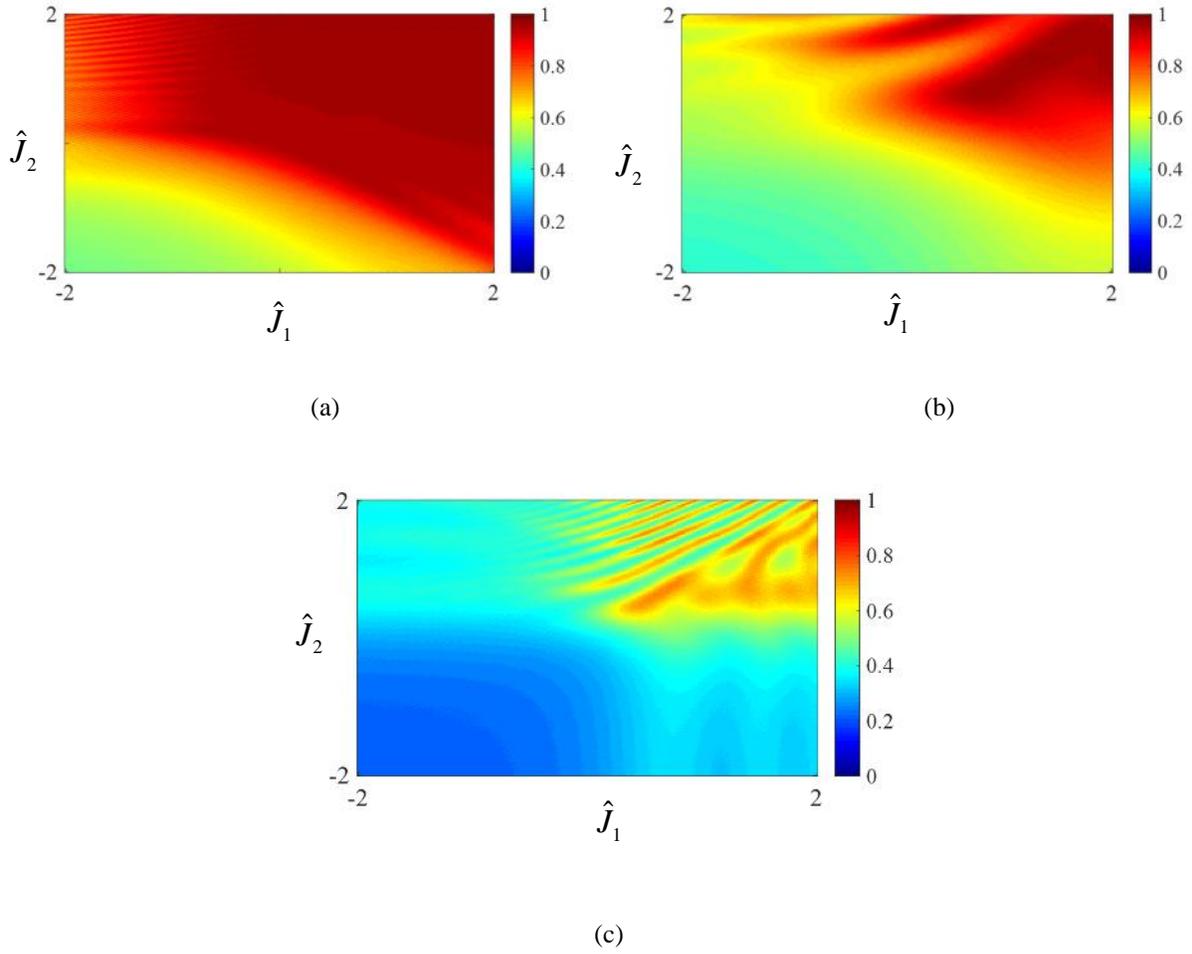
Fig 5: FTPE for a four particle system while second neighbor interactions are included while (a): $\hat{r} = 1$, (b) $\hat{r} = 7$ and (c): $\hat{r} = 13.8$.